\documentclass[a4paper,11pt]{article}
\pdfoutput=1 

\usepackage{jheppub} 

\usepackage[T1]{fontenc} 
\usepackage{amsmath}
\usepackage{amsfonts}
\usepackage{amsmath}
\usepackage{amsxtra}
\usepackage{amstext}
\usepackage{amssymb}
\usepackage{slashed}
\usepackage{graphicx}
\usepackage{cleveref}
\usepackage{color}

\numberwithin{equation}{section}

\newcommand{\beq}{\begin{equation}}
\newcommand{\eeq}{\end{equation}}
\newcommand{\bea}{\begin{eqnarray}}
\newcommand{\eea}{\end{eqnarray}}

\newcommand{\vep}{\varepsilon}
\newcommand{\ep}{\epsilon}

\newcommand{\vphi}{\varphi}

\newcommand{\der}{\partial}
\newcommand{\derb}{\overline{\partial}}

\newcommand{\<}{\langle}
\renewcommand{\>}{\rangle}
\newcommand{\N}{{\cal{N}}}
\newcommand{\im}{\mathrm{Im}\,}
\newcommand{\re}{\mathrm{Re}\,}
\newcommand{\z}{\zeta}

\DeclareMathOperator{\Tr}{Tr}

\title{\boldmath K\"ahler moduli stabilization from ten dimensions}

\author[a]{Iosif Bena,}
\author[a]{Mariana Gra\~na,}
\author[b]{Nicolas Kovensky,}
\author[a]{Ander Retolaza.}

\affiliation[a]{Institut de Physique Th\'eorique, Universit\'e Paris Saclay, CEA, CNRS, Orme des Merisiers, 91191 Gif-sur-Yvette CEDEX, France.}
\affiliation[b]{Mathematical Sciences and STAG Research Centre, University of Southampton, Southampton, SO17 1BJ, United Kingdom.}

\emailAdd{iosif.bena@ipht.fr}
\emailAdd{mariana.grana@ipht.fr}
\emailAdd{n.kovensky@soton.ac.uk}
\emailAdd{ander.retolaza@ipht.fr}

\abstract{We describe the back-reaction of gaugino condensates in supersymmetric AdS$_4$ Type II String Theory compactifications with fluxes. We use generalized complex geometry to capture the modification of the ten-dimensional supersymmetry equations and show that the cosmological constant prevents the cycle wrapped by the branes with gaugino condensation from shrinking to zero size. Thus, unlike in ordinary geometric transitions in flat space, the volume of this cycle remains finite. For D7 branes with gaugino condensation, this gives a ten-dimensional account of K\"ahler moduli stabilization. Furthermore, by    matching the ten-dimensional supergravity solutions near and far from the cycle wrapped by the D7 branes, we find a   relation between the size of this cycle and the  cosmological constant. This relation agrees with the supersymmetric AdS vacuum condition obtained by KKLT using effective field theory.
}

\begin{document} 
\maketitle
\flushbottom

\section{Introduction}
\label{sec:intro}

Non-perturbative contributions to a Lagrangian can typically be safely ignored in the weakly coupled regime, as they are completely sub-leading with respect to tree-level and perturbative terms.   However, whenever a given coupling is absent at the perturbative level, the non-perturbative terms are no longer a correction and can completely change the physics. This is what happens for compactifications of type IIB string theory on Calabi-Yau manifolds with fluxes, where the fluxes do not couple to the moduli parametrizing the deformations of the K\"ahler form of the manifold. This is particularly problematic since the overall volume of the manifold is a combination of K\"ahler moduli (or is {\em the} K\"ahler modulus if there is only one), and thus the absence of couplings implies that any volume is allowed. 

Kachru, Kallosh, Linde and Trivedi (KKLT) \cite{Kachru:2003aw} realized that non-perturbative effects, arising either from gaugino condensation on D7-branes or from Euclidean D3-brane instantons, can give rise to the desired couplings, generating a potential that stabilizes these moduli and giving rise to a supersymmetric AdS$_4$ warped compactification. Furthermore, they proposed a mechanism to uplift the negative cosmological constant of these compactifications to positive values, making them interesting for explaining the current accelerated expansion of our Universe.

In the KKLT proposal, the important physics happens at three well-separated  scales, which allows one to split the  construction into three steps. Each step addresses one concrete  physical  phenomenon while using an effective field theory that captures  the main features of the previous step:

\begin{itemize}
    \item The starting point is a Type IIB solution \`a la Giddings-Kachru-Polchinski \cite{Giddings:2001yu,Grana:2000jj}, with imaginary self-dual 3-form fluxes on an conformally-Calabi-Yau internal manifold and a vanishing four-dimensional cosmological constant. At this stage, the complex structure moduli and the dilaton get fixed by the 3-form fluxes. However, this kind of no-scale compactification leaves the K\"{a}hler moduli unfixed.    
    
    \item One then includes non-perturbative (NP) effects, coming either from gaugino condensation on D7-branes or from Euclidean D3-branes instantons wrapping four-cycles of the internal manifold. Such  effects generate a NP contribution to the 4D superpotential that depends on the K\"ahler moduli, and in the effective field theory for these moduli one finds a supersymmetric AdS$_4$ solution whose negative   cosmological constant comes from the non-zero on-shell value of the superpotential, $W$ 
    \footnote{Another proposal to stabilize K\"ahler moduli is to use NP effects together with the first $\alpha '$ correction  \cite{Cicoli:2007xp}. Unlike in KKLT, this mechanism leads to non-supersymmetric AdS vacua. In this paper we use the supersymmetry conditions and stay at leading order in $\alpha '$, so we do not consider this type of vacua. }.
    
    \item One finally adds a small number of anti D3-branes at the tip of a long warped throat. These must be located far from the D7-branes, and the redshift is argued to guarantee a negligible effect on bulk physics, providing control on the backreaction on moduli stabilization\footnote{In this paper we restrict ourselves to supersymmetric configurations. Nevertheless, there have been concerns regarding the  uplifting mechanism \cite{Bena:2009xk,Blaback:2011pn,Bena:2014jaa,Moritz:2017xto,Gautason:2018gln,Bena:2018fqc}. Moreover,  the difficulties in obtaining solutions with a positive cosmological constant  have  led to the conjecture that all de Sitter constructions in string theory that can be described using effective field theory are not stable. This is known as the de Sitter Swampland Conjecture \cite{dSswampland}. }.     These break supersymmetry and induce an energy contribution suitable for uplifting the cosmological constant to values even above zero. In principle, this has the advantage of producing a parametrically small $\Lambda$.
\end{itemize}

Although very appealing, up to date there is an on-going debate on whether this proposal is consistent. While the first step is well understood\footnote{Though already at the first step, there are questions regarding the validity of the supersymmetry-breaking Minkowski solution with non-zero $W$ \cite{Sethi:2017phn}.}, it is not clear if the other two steps involved in this construction stand on firm ground. In this paper we focus on the first and second steps, making only some brief comments on the third one. 

\medskip

While the first step is done at the 10D level, to account for NP effects one needs to resort to an effective 4D theory. Integrating out the complex-structure moduli and dilaton which are supposed to be fixed at a higher energy\footnote{The use of  effective field theory including warped modes  was recently shown to be much more subtle than it appears,  as the complex-structure modulus corresponding to the size of the $S^3$ has also an exponentially small mass \cite{Bena:2018fqc, Blumenhagen:2019qcg}.}, one gets a non-perturbative superpotential for the K\"ahler moduli, whose imaginary parts are the volumes of 4-cycles. This is based on the relation of these moduli with the (imaginary part of the holomorphic) gauge coupling on the D7-branes, $4 \pi g_{\mathrm{YM}}^{-2} = \frac{1}{g_s}\sigma$, where $\sigma$ is the size of the 4-cycle in string units. If the chiral fields on the D7-branes are massive, the low-energy physics on the world-volume theory, as seen from four-dimensions, is described by a pure $\N=1$ Super Yang-Mills theory, which undergoes gaugino condensation. The associated non-perturbative superpotential and the strength of the condensate are related to $\sigma=\im T$ by 

\begin{equation}
    W_{\mathrm{NP}} \sim \<\lambda\lambda\> = A \exp \left({i a T} 
    \right), \quad \
    \label{WnpKKLT}
\end{equation}
for some constant $A$\footnote{More precisely, $A$ depends on the complex-structure and open-string moduli, but at this stage these are taken to be fixed.}. For $SU(N_{D_7})$ one has $a=\frac{2\pi}{N_{D7}}$ \footnote{The functional dependence of this constant is different for other gauge groups, such as $SO/USp$ (which arise when the branes are on top of orientifold planes). }.   In the simplest compactifications with a single K\"ahler modulus (and hence a single NP term), the K\"{a}hler potential is $K(T) =  -3 \log (-i(T-\bar T))=-3 \log (2\sigma)$. The non-perturbative effect  balances against some classical and, in principle, tunable contribution $W_0$ coming from the fluxes, leading to the following F-term condition on the supersymmetric solution\footnote{Note that $W_0$ has to be very small in order to compete with the NP effect and stabilize the  K\"{a}hler modulus. This smallness and the exponential dependence of the NP effect on the modulus lead to a large cycle size, which allows one to control higher-order NP corrections.   Nevertheless, it is still not clear that the fluxes can be tuned to make $W_0$ so small.}:
\begin{equation} \label{KKLTminumum}
 W_0 = -A e^{-a \sigma_*} \left(1+\frac{2}{3} a \sigma_*\right)\quad \rightarrow \quad   \Lambda = -3e^{-K( \sigma_* )}  |W_{\mathrm{on-shell}}|^2 = - \frac{a^2 A^2 e^{-2a\sigma_*}}{6 \sigma_*} \ , 
\end{equation}
When the non-perturbative effects come from Euclidean D3-branes wrapping the same 4-cycle $\Sigma_4$, the analogous F-term condition is obtained by replacing \ $N_{\text{D7}} \rightarrow 1$  \  \cite{Witten:1996bn}. 

Even if this second step involves reasonably well-known ingredients in string theory, the existence and main features of the supersymmetric vacuum solution  were obtained  using 4D supergravity. This raises the question of whether the effective field theory approach captures the relevant physics (namely, the vacuum solution as well as its deformations), or if the higher dimensional analysis may reveal new features of the compatification that the effective field theory misses. In particular, the inclusion of NP effects leads to a small 4D (negative) cosmological constant, resulting in an  AdS scale much larger than the KK scale, a feature that has recently been conjectured not to be possible in the landscape \cite{Gautason:2018gln,Lust:2019zwm}. On the other hand, 
the AdS background is the seed solution to ultimately obtain de Sitter compactifications, and in this process the NP effects are key in evading classical no-go theorems \cite{deWit:1986mwo,Maldacena:2000mw}.
Recent results revealed that in order to properly address whether the KKLT uplifting mechanism can work or not using a Maldacena-N\' u\~{n}ez-like approach, 
 it is crucial to understand what is the effect of NP physics on  the geometry \cite{Moritz:2017xto,Gautason:2018gln,Hamada:2018qef,Kallosh:2019oxv,Hamada:2019ack,Gautason:2019jwq,Carta:2019rhx}.

 With all these motivations in mind, our purpose is to present the main features of the supersymmetric AdS vacuum and the non-perturbative effects that give rise to it, directly in the ten-dimensional geometry. Let us first comment on some crucial issues: first, it is not clear that gaugino condensation actually happens when the extended part of the D7 brane worldvolume is not flat. Here we will assume it does, without studying the whole dynamical process leading to the condensate developing a vev\footnote{Alternatively, we could consider NP contributions sourced by Euclidean D3-brane instantons wrapped on the internal space.}.  A second non-trivial issue is whether these effects can be captured by ten-dimensional supergravity. Based on previous work, and on the results of this paper, we will argue they can. Hence, the main questions about the 10D description of the supersymmetric $\N=1$ AdS$_4$ vacua are: 

\begin{itemize}
    \item How do we include the NP effects in the 10D supersymmetry conditions?
    \item How do these effects back-react on the internal geometry and fluxes? 
\end{itemize}

The first step towards answering the first of these questions was beautifully addressed by Koerber and Martucci in \cite{Koerber:2007xk,Koerber:2008sx}, 
using the framework of Generalized Complex Geometry (GCG; for a review, see \cite{Koerber:2010bx}), which we shall use. 
They first showed (see also \cite{Cassani:2007pq}) that the conditions for ${\cal N}=1$ supersymmetry (in the absence of condensates) \cite{Grana:2005sn} 
 are equivalent to the F- and D-flatness conditions for the 10D off-shell superpotential \cite{Grana:2005ny,Grana:2006hr,Benmachiche:2006df} that generalizes the Gukov-Vafa-Witten  superpotential \cite{Gukov:1999ya} for 3-form fluxes on Calabi-Yau manifolds.  Then they were able to identify how some of these conditions are modified in the presence of non-perturbative contributions from space-filling D7 branes wrapping four-cycles on the internal manifold. One of the most important observations is that the back-reaction modifies the structure of the manifold. Thus, even if one started from a flux compactification on a Calabi-Yau manifold, or more generally on a manifold of SU(3) structure (but not necessarily of SU(3) holonomy), the gaugino condensate on the D7 branes deforms the structure into a more general (pair of) generalized complex structures.
 
The second question was studied later in several papers, from different points of view. A computation of the first-order back-reaction felt by probe D3-branes was presented in \cite{Baumann:2010sx} in the context of the AdS/CFT correspondence, giving a \textit{geometrization} of the non-perturbative effects in terms of imaginary anti-self-dual (IASD) fluxes. These fluxes were proposed to be sourced via a coupling (first studied by \cite{Camara:2004jj} for other purposes) in the fermionic D7-brane action between the gaugino mass term and the IASD fluxes. A more precise analysis of the back-reaction    was later performed in \cite{Dymarsky:2010mf} for a non-compact transverse manifold. Finally,  a detailed study of one particular example was carried out in \cite{Heidenreich:2010ad}, which proposed to use a dynamic $SU(2)$ structure to characterize the generalized deformation.

This paper has three purposes. The first is to connect the  approaches above. The second is to find the 10D mechanism by which the K\"ahler moduli are stabilized. We show that in the presence of a 4D cosmological constant, one of the supersymmetry equations, \eqref{susy1b}, implies that the four-cycle wrapped by D7 branes with a non-trivial gaugino condensate cannot collapse.  Hence these D7 branes do not undergo the usual geometric transition. The third purpose is to use this 10D description to obtain a relation between the parameters of the vacuum solution and see if it can match the EFT results. We  find that the size of the four-cycle is determined by matching the solution near the D7 branes (in the region where the non-perturbative effects sourced by the gaugino condensate dominate over the 4D cosmological constant), and the solution far from them (where the cosmological constant dominates). This gives us equation \eqref{W0us}, which confirms the validity of the KKLT relation \eqref{KKLTminumum} obtained using four-dimensional EFT.

As we will explain in section \ref{sec:condensate}, our results also imply that a 4D cosmological constant is incompatible with D5-brane Killing spinors, even in the presence of non-perturbative effects. In particular, backgrounds that contain geometrically-transitioned D5 branes and have D5 brane Killing spinors, such as the Maldacena-N\'u\~nez throat \cite{Chamseddine:1997nm, Maldacena:2000yy} or the Baryonic-Branch throat \cite{Butti:2004pk} cannot be part of a supersymmetric AdS compactification. Note that this is a local constraint, and hence it is much more powerful than the global arguments that have been used in the past against the presence of baryonic-branch throats in flux compactifications.

The paper is organized as follows. We begin in Section \ref{sec:GG} by very briefly reviewing compactifications in the language of GCG, including some simple examples.  In Section \ref{sec:condensate} we review previous results on the 10 dimensional description of non perturbative effects in the framework of GCG,  and combine them to show how moduli stabilization happens from a 10D point of view. Then, in Section \ref{sec:SU2} we focus on the solution in \cite{Heidenreich:2010ad} and describe how the matching of the near- and far-brane behaviors determines the 10D equation governing the volume stabilization and the vacuum KKLT solution. Finally, we discuss the consequences of our results and the possibilities for future lines of work in Section \ref{sec:conclusions}.

\section{Generalized geometry language for type II compactifications}
\label{sec:GG}

Consider type II superstring theory on a warped product of a four-dimensional Mink$_4$ or AdS$_4$ manifold, and a compact internal six-dimensional manifold $M_6$:
\beq \label{10Dmetric}
ds_{10}^2=e^{2A(y)} g_{\mu\nu}(x) dx^\mu dx^\nu + h_{mn}(y) dy^m dy^n \ ,
\eeq
where $x^\mu, \mu=0,..3$ are coordinates on Mink$_4$ or AdS$_4$, and $y^m$, $m=1,..6$ are coordinates on $M_6$.
The most general 10D Majorana-Weyl supersymmetry spinors of solutions that preserve $\N=1$ supersymmetry in four dimensions split as
 \begin{equation}
    \ep_1 = \zeta_+ \otimes \eta^1_+ + \zeta_- \otimes \eta^1_- \ , \ 
    \ep_2 = \zeta_+ \otimes \eta^2_\mp + \zeta_- \otimes \eta^2_\pm \ 
\end{equation}
where $\z_+^*=\z_-$ and $\eta_+^{i*}=\eta_{-}^i$. Although most of the analysis of this paper takes place in  type  IIB string theory, some results apply equally well to  type IIA compactifications, so we will use conventions in which the upper (lower) sign corresponds to type IIA (IIB). The four-dimensional spinors satisfy
\beq
2\nabla_\nu \zeta_- = \pm \mu \gamma_{\nu} \zeta_+
\eeq
with $\mu$ the value of the on-shell superpotential, such that the 4D cosmological constant is $\Lambda = -3|\mu|^2$. The internal spinors $\eta^{i}_+$ are globally defined and identically normalized. 

Using these spinors one can build two polyforms or pure spinors which characterize the background geometry, and are defined as
\begin{equation} \label{purespinors}
    \Psi_{\pm} \equiv -\frac{8 i}{||\eta||^2} \sum_p \frac{1}{p!} \eta^{2\dagger}_{\pm} \Gamma_{m_1 \dots m_p} \eta^1_+ \, dy^{m_1} \wedge \dots \wedge dy^{m_p} \ . 
\end{equation}
These spinors define an $SU(3) \times SU(3) \subset \rm{Spin}(6,6)$ structure, and contain all the information about the internal metric. 

We introduce for later use the Mukai pairing, which is the form version of the inner product between Spin(6,6) spinors
\begin{equation} \label{Mukai}
    \<\Psi,\Phi\> \equiv \Psi \wedge \alpha(\Phi)|_{\mathrm{top}} \ , \quad  \alpha(\omega_p)=(-1)^{\frac{p(p-1)}{2}}\omega_p. 
\end{equation}
where $|_{\mathrm{top}}$ means the top-form piece (for compactifications to 4 dimensions this is a 6-form).

As found in \cite{Grana:2005sn}, the pure spinors allow a compact and elegant rewriting of the supersymmetry conditions $\delta \psi_M = \delta \lambda=0$  in the string frame:
\begin{subequations}
\label{susy}
 \begin{align}
d_H \left(e^{3A-\phi} \Psi_2\right) &= 2 i \mu e^{2A-\phi} \im \Psi_1 \label{susy1}\\
d_H \left(e^{2A-\phi} \im \Psi_1 \right) &= 0 \label{susy2}\\
d_H \left(e^{4A-\phi} \re \Psi_1 \right) &= 3 e^{3A-\phi} \re (\bar{\mu}\Psi_2) + e^{4A} *_6 \alpha(F) \label{susy3}
 \end{align}
\end{subequations}
where $F$ is the sum of the internal RR fluxes, 
\beq
\Psi_1 = \Psi_\mp \ , \qquad \Psi_2 = \Psi_\pm \quad {\rm upper \ (lower) \ sign \ for \ IIA \ (IIB)}
\eeq
and we have used the $H$-twisted exterior derivative $d_H \equiv d + H_3\wedge$. Of course, for $\mu$ non-zero \eqref{susy2} is a consequence of \eqref{susy1}.  For $\mu=0$ the first two of these equations  are   twisted integrability conditions. Once an internal generalized geometry satisfying these equations is specified, one can use  the last equation to fix the  RR fluxes necessary to build a consistent supersymmetric ten-dimensional supergravity background. The equations of motion (EOMs) of these fluxes are automatic, but one still has to impose by hand the EOM of the NSNS flux $H_3$ and all the Bianchi identities; the latter can be written schematically as 
\begin{equation}
    d_H \, F = \mathrm{sources} \ . 
\end{equation}

We quickly review the familiar type IIB  compactifications on $SU(3)$ structure manifolds with $\mu=0$. These only have one well-defined spinor on the internal space and therefore the two spinors on the decomposition of 10D supersymmetry parameters  differ at most by a relative phase.  We take $\eta^2_+=-ie^{i\theta} \eta^1_+$, such that $\theta=0$ for D3/D7-type supersymmetry, while $\theta=\pi/2$ for D5-type supersymmetry. The pure spinors are
\begin{equation}
    \Psi_2 =\Psi_-= i e^{-i \theta} \Omega \ , \ \Psi_1 = \Psi_+=e^{-i \theta} \exp (-i J), 
\end{equation}
where $J$ and $\Omega$ are the  real (1,1)-form and holomorphic (3,0) form defining the SU(3) structure. 
For D3/D7-type supersymmetry, $\theta=0$, and the conditions \eqref{susy2}, \eqref{susy1} imply respectively that 
$e^{2A-\phi} J$ (the 2-form in $\im \Psi_1$) and $e^{3A-\phi} \Omega$ are closed. For constant dilaton\footnote{The equations impose  the axion-dilaton $\tau \equiv C_0 + ie^{-\phi} $ to be holomorphic, allowing for D7-brane configurations. Constant dilaton is a particular subcase.}, these imply that the metric $e^{2A} h_{mn}$ is Calabi-Yau. Eq. \eqref{susy3}, the six-form in \eqref{susy1} and the five-form in \eqref{susy2} say respectively that $e^{-\phi} H_3 = -*_6 F_3$, $H_3\wedge \Omega=0$ and $J \wedge H_3=0$. Equivalently,  the complex 3-form flux 
\beq \label{G3}
 G_3 \equiv F_3 + i e^{-\phi} H_3 \, 
 \eeq
 is imaginary self-dual (ISD), primitive and has no (0,3) component (in other words, it is (2,1)). The one-form component of \eqref{susy3} relates the warp factor and $F_5$ by $d(4A-\phi) = e^{\phi} *_6 F_5$. For backgrounds with calibrated D5-branes $\theta=\pi/2$ and the real and imaginary parts of $\Psi_1$ get switched, which implies, among other things, that $F_3$ is related to $dJ$ (instead of   $H_3$) through an ISD condition (we expand on this later, in \eqref{Ggen})

In type IIA, with the supersymmetry of D6-branes, one sees that the two pure spinors exchange roles. This is nothing but the mirror-symmetric picture of the type IIB construction.

We will see that $SU(3)$ structure is too restrictive to account for the back-reaction of D7-branes with gaugino condensates, where the angle  between the two supersymmetry spinors depends on the distance to the brane (this is not the phase of spinors of solutions with an SU(3)-structure, but an angle between the two well defined spinors of these more general solutions). In terms of a structure group on the internal manifold, this is called ``dynamic SU(2) structure'' (as opposed to a ``rigid SU(2)'' where the two spinors are always orthogonal); however, as a structure group on the tangent bundle it is not well defined since the spinors become parallel  at some points, this $SU(2)$ structure is not globally defined. This is where GCG  comes at play: all of these structures become the same from the point of view of the generalized tangent bundle, they are all $SU(3)\times SU(3)$ structures, fully encoded in the pure spinors $\Psi_+, \Psi_-$ in \eqref{purespinors}. We will come back to this point in section \ref{sec:SU2}, where we will give the polyform expression for the pure spinors for a dynamic SU(2) structure. 

\subsection*{Ten-dimensional perspective}

One can show that eqs. \eqref{susy} can be derived from a superpotential. Going back to the CY example, we know that some of these equations, more precisely those related with the 3-form flux $G_3$ in \eqref{G3}, can be understood by looking at the Gukov-Vafa-Witten (GVW) superpotential \cite{Gukov:1999ya} (we work with $l_s=1$)
\begin{equation}
    W_{\mathrm{GVW}} = -   
    \int_{M_6} \Omega\wedge G_3,
    \label{GVW}
\end{equation}
which was obtained by studying domain-wall tensions in the four-dimensional effective theory. In this truncated effective theory the fluctuations only include the massless modes on the CY, and there is no warping. 
Thus, this is not suited for deriving the complete ten dimensional supersymmetry conditions in more general compactifications.

This puzzle was beautifully solved in a series of papers \cite{Grana:2005ny,Grana:2006hr,Benmachiche:2006df,Koerber:2007xk,Koerber:2008sx}. Keeping in mind the familiar expressions in backgrounds compatible with D5 and  D6-branes
 \begin{equation} \label{D5,D6}
W_{\rm D5} = -   
\int_{M_6} \Omega\wedge \left(
F_3 + i e^{-\phi} dJ\right) \ , \ 
W_{\rm D6} =   
\int_{M_6} (J - i B) \wedge \left(
F_4+i d[e^{-\phi}\mathrm{Re}\,\Omega]\right),   
\end{equation}
the authors showed that the form of the 10D off-shell superpotential for warped compactifications in terms of the pure spinors and the Mukai pairing is given by 
\begin{equation}
    W_{10\rm D} =  
    \int_{M_6} \<e^{3A-\phi} \Psi_2,  d_H [C + i e^{-\phi} \mathrm{Re} \Psi_1]\> =   \int_{M_6} \<Z, d \, T\>    
    \label{W10D}
\end{equation}
where $C$ are the RR gauge potentials ($d_H C = F $) and in the last equality we have introduced the proper holomorphic fields in 10D\footnote{Note that in the last equality in \eqref{W10D} we switched from the twisted exterior derivative to the ordinary one, and twisted the pure spinors instead by $e^B$. Both formulations are equivalent but the latter is necessary to define the proper holomorphic fields which include the B-field.}  
\beq \label{Z,T}
Z = e^{3A-\phi} e^B \Psi_2 \ , \qquad T=e^B ( C + i e^{-\phi} \mathrm{Re}\,  \Psi_1 ) \ . 
\eeq
 
Given this superpotential, Refs \cite{Koerber:2007xk,Koerber:2008sx} and \cite{Cassani:2007pq} showed that the F-flatness and D-flatness
conditions amount to the full  supersymmetry conditions \eqref{susy}. 
For that, one has to perform a dimensional reduction, using the (string-frame) metric Ansatz \eqref{10Dmetric}. Crucially, the reduction is done without the need to specify a particular set of modes to expand the fields into (such as the harmonic forms used in CY compactifications). The first supersymmetry equation, \eqref{susy1}, which is the one we will focus on, was shown to be  equivalent to the F-flatness condition for the modulus $T$.
This is not hard to see for $\mu=0$, by integrating $W_{10D}$ in \eqref{W10D} by parts. When the cosmological constant is not zero, one has to take the covariant derivative of the superpotential, and the extra term of this covariant derivative (involving the superpotential times the derivative of the K\"ahler potential) gives precisely  the right-hand side of \eqref{susy1}.

\section{Gaugino condensation}
\label{sec:condensate}

As discussed in the introduction, one of the necessary ingredients of the KKLT proposal is the inclusion of NP effects. We want to see how to include the NP effects in the supersymmetry conditions, and how they backreact on the internal geometry and fluxes. We concentrate on the situation where the NP contribution is coming from gaugino condensation induced by strong-coupling physics of the $\N=1$ SYM theory on $N_{D7}>1$ space-time filling D7-branes wrapping an internal four-cycle $\Sigma_4$.

Before trying to answer this question, let us make a few remarks. First, if one wants $M_6$ to be compact it is mandatory to ensure that all $p$-form charges add-up to zero. This implies that one must also have O7-planes in the background. In what follows, we do not need to assume anything about the location of O7-planes, they could be on top of the D7-branes (changing the gauge group from $SU(N_{D_7})$ to $SO(2N_{D7})$) or further apart.  Second, it is important to recall that supersymmetry forces $\Sigma_4$ to be a generalized calibrated manifold\footnote{For a review and a more complete set of references, see \cite{Koerber:2010bx}.}  \cite{Koerber:2005qi,Martucci:2005ht}. In the absence of world-volume fluxes ${\cal{F}}$ and for a Calabi-Yau manifold (or, more generally, a manifold with an SU(3) structure) this implies that $\Sigma_4$ has to be a complex sub-manifold, whose volume form  is given by $1/2 \, J^2|_{\Sigma_4}$. 

For more general solutions (with an SU$(3)\times$SU$(3)$ structure and/or world-volume fluxes), the (algebraic part of the) calibration condition for a supersymmetric space-filling D$p$-brane in a type II vacuum can be written in the GCG language as the algebraic condition\footnote{An analogous statement can be derived for Euclidean branes wrapping   $\Sigma$, see  Appendix D of \cite{Koerber:2007xk} or \cite{Koerber:2008sx}.} 
\begin{equation}
    e^{4A-\phi} \sqrt{\det{\left[(g+B)|_\Sigma+{\cal{F}}\right]}} d\xi^1 \wedge ... d\xi^{p-3} \,= e^{4A-\phi} \re \Psi_1|_\Sigma \wedge e^{{\cal{F}}} |_{p-3}
\end{equation}
where $\xi$ are the world-volume coordinates. It can be seen \cite{Koerber:2005qi,Koerber:2007jb}  that the differential condition leading to the volume minimization of the cycle is nothing else but the   supersymmetry equation \eqref{susy3}. Interestingly, the rest of the conditions \eqref{susy} can also be interpreted in a similar way. Indeed, $e^{3A-\phi}\Psi_2$ and $e^{2A-\phi} \im \Psi_1$ constitute the calibration forms corresponding to domain walls and string-like objects, respectively.     

Going back to the condensate, we need to calculate how its presence affects the geometry and the fluxes. In particular, it is crucial to understand  the modifications it induces on the supersymmetry conditions \eqref{susy}. Here we follow an instructive argument presented in \cite{Dymarsky:2010mf}. 
 We are looking for the modification of the 10D supersymmetry equations associated to  the NP contribution \eqref{WnpKKLT} in the superpotential\footnote{This contribution can be computed explicitly  in Heterotic theory, where the four-fermion terms in the 10D action are known. Intriguingly, it was found in  \cite{Frey:2005zz} that there is no fermion bilinear contribution to the 4d superpotential. Here we follow KKLT and assume that such a contribution does exist. It would be interesting to understand the relation between the Heterotic and the Type II results. }. As described above, this term depends on the gauge coupling that in the 10D picture is given by the volume of the 4-cycle $\Sigma$, hence the dependence of \eqref{WnpKKLT} on the imaginary part of the K\"{a}hler modulus $T$. The ${\cal F}^2$ part of the DBI action for the Dp-branes is
 \begin{equation}
    S_{Dp}|_{{\cal{F}}^2} = -\frac{1}{8\pi 
    }\int_\Sigma e^{-\phi} \re \Psi_1 \int d^4x \sqrt{-g} \Tr {\cal{F}}^2 \ , 
\end{equation}
from where we one can read off the effective SYM coupling constant.
Furthermore, we can easily guess the holomorphic completion of this coupling: $e^{-\phi} \re \Psi_1$ will be replaced by the (untwisted) variable $T$ defined in \eqref{Z,T}. Indeed, the effective SYM $\theta$ term comes from the Wess-Zumino term in $S_{Dp}$, where the world-volume fluxes are coupled to the RR potentials. Finally, note that if we define $\delta^{9-p}[\Sigma]$ as the $(9-p)$-form Poincar\'e dual  to the $(p-3)$-cycle $\Sigma$ wrapped by the Dp-brane, such that 
\begin{equation}
    \int_{M_6} \omega \wedge \delta^{9-p}[\Sigma] = \int_{\Sigma} \omega|_\Sigma 
\end{equation}
for any ($p-3$)-form $\omega$, we can write the 4D holomorphic coupling as
\begin{equation} \label{eq:local-gauge}
    \tau = 
    \int_\Sigma ( C+ i e^{-\phi} \re \Psi_1 )|_\Sigma = 
    \int_{M_6} T \wedge \delta^{9-p}[\Sigma] = 
    \int_{M_6} \<T ,\delta_\alpha^{9-p}[\Sigma]\>.  
\end{equation}
with $\delta_\alpha^{9-p}[\Sigma] \equiv \alpha (\delta^{9-p}[\Sigma])$ and $\alpha$  defined in \eqref{Mukai}. 
We are now ready to present how the non-perturbative effects modify the first supersymmetry equation \eqref{susy1} which, we recall, is equivalent to the F-flatness condition for the modulus $T$.   Taking the covariant variation of the full   superpotential (namely the sum of the non-perturbative piece \eqref{WnpKKLT} and the perturbative one \eqref{W10D}) with respect to the 10D holomorphic variable $T$ one concludes that \eqref{susy1} gets an additional non-perturbative term. The exponential dependence of the NP term on $T$ makes this new 10D contribution proportional to the  4D vev. Moreover, since the gauge kinetic function  \eqref{eq:local-gauge} is localized, this new term is localized. Hence, the quantum corrected version of  \eqref{susy1} is  
\begin{equation}
    d_H \left(e^{3A-\phi} \Psi_2\right) = 2 i \mu e^{2A-\phi} \im \Psi_1 + 2 i 
    \<S\> \delta_\alpha^{9-p} [\Sigma]. \label{susy1b}
\end{equation}
where $S$ is the usual superfield associated with the gaugino condensate with expectation value 
\beq
 \<S\> = \frac{1}{16\pi^2} \<\lambda \lambda\>.
 \eeq
Finally, $\mu$ is still related to the on-shell value of the superpotential, which is also modified in order to include the NP term: $W_{\mathrm{10D}} \to W_{\mathrm{10D}} + W_{\mathrm{NP}}$.

\subsection*{Interpretation}

Now that we know how \eqref{susy1} is modified, we would like to understand its implications, focusing on some examples.  We are interested in the local features of the geometry driven by the new term in \eqref{susy1b}, so we will leave aside extra  constraints arising in compactifications, such as tadpole cancellation, and make our conclusions based on results obtained in non-compact geometries.

For $\mu = 0$, equation \eqref{susy1b} implies that  $d_H Z \sim  \<S\> \delta^{9-p} [\Sigma]$. This has a very simple geometrical interpretation \cite{Garcia-Valdecasas:2016voz,Tenreiro:2017fon}: it tells us that $\delta^{9-p} [\Sigma]$ is trivial in the cohomological sense\footnote{Note that the differential operator is the twisted exterior derivative, and thus the dual cycles are trivial in twisted cohomology. However, for D7-branes the right hand side is a two-form, and therefore the twisting by $H_3$ plays no role and the dual four-cycle would be trivial in ordinary cohomology.}. Stokes' theorem then implies that integrating any closed ($p-3$)$-$form on  $\Sigma$ will give zero. In other words, the cycle itself shrinks and becomes trivial in homology.  This is familiar in the context of  geometric transitions. For example, for D5 branes wrapping the holomorphic 2-cycle of at the tip of a resolved conifold, \eqref{susy1b} implies that $d \Omega_3 \sim \<S\> \delta^{4} [\Sigma_2]$. The strong-coupling dynamics make $\Sigma_2$ shrink, while the 3-cycle blows up and becomes topologically non-trivial, regularizing the geometry into a deformed conifold. The D5-branes disappear and the final solution has no localized sources. Furthermore, it can still be described as a manifold with $SU(3)$-structure.
 By using the AdS/CFT correspondence, this transition admits a very nice interpretation from the CFT point of view: it encodes the chiral symmetry breaking that takes place in the IR of the theory. Of course, a mirror picture works for D6 branes, where one has\footnote{Here $\Sigma_3$ needs to be a special Lagrangian cycle for the D6-branes to be calibrated. Moreover,  the $H_3$ twist on the exterior derivative is irrelevant  for us in this class of Type IIA solutions because Re $\Psi_2$ has a 2-form component but no 0-form component. } $d J \sim \<S\> \delta^{3} [\Sigma_3]$. This was first studied in \cite{Vafa:2000wi} (see also \cite{Atiyah:2000zz}). 

Now, as emphasized in \cite{Dymarsky:2010mf}, the situation with D7-branes is very different from the previous examples. First, a gauge theory dual  for D7-branes wrapping holomorphic 4-cycles is difficult to find. In particular  one cannot take $N_{\mathrm{D7}}$ large. Concerning the supergravity solution we are analyzing, there are two fundamental differences for D7-branes: 
 
\begin{itemize}
   
    \item For $p=7$ the cycle $\Sigma_4$ is a holomorphic 4-cycle and the localized delta function is a 2-form. Thus,  Eq.\eqref{susy1b} has a term $d \left(e^{3A-\phi} \Psi_2\right) \sim \<S\> \delta^{2} [\Sigma]$ and so $\Psi_2$ necessarily contains a 1-form. As a consequence, it is impossible to satisfy the supersymmetry condition while staying within the realm of ordinary complex geometry \cite{Dymarsky:2010mf}. We are forced to consider manifolds with more general structure groups than SU(3). The generic structure group is the so-called  dynamical $SU(2)$ structure.

    \item Moreover, something very interesting happens if we consider an AdS$_4$ compactification.  For $p=7$ and $\mu \neq 0$ Eq. \eqref{susy1b} takes the form 
        \begin{equation}
         d_H \left(e^{3A-\phi} \Psi_2\right) = 2 i \mu e^{2A-\phi} \im \Psi_1 - 2 i
         \<S\> \delta^{2} [\Sigma_4].
         \label{susy1bD7}
    \end{equation}
     The first observation we make is that, since $\mu$ is nonzero, $\delta^2[\Sigma_4]$ is not exact anymore.   Following the logic in the above arguments, this means that the cycle wrapped by the D7-branes is not  trivialized anymore! The situation is different from the usual geometric transitions. Moreover, it is consistent with the interpretation we would like to give to this configuration in the context of the KKLT scenario. Indeed, the stabilization of the volume modulus can be understood from \eqref{susy1bD7} in terms of $\Sigma_4$ staying at a definite, finite size\footnote{Note that in order for the gauge coupling to still be large such that the gauginos get a vev it is necessary that the cycle is stabilized at a small size. } and the presence of a negative cosmological constant in four dimensions constitutes a necessary condition. To the best of our knowledge, this interpretation has not been given in the literature before.  
     
\end{itemize}

Because we were interested in local features, we decided to leave aside global constrains. Nevertheless, note that in a global compactification, \eqref{susy1b} provides a relation between the amplitudes of  all NP effects and the 4D CC in the form of a tadpole-cancellation condition.  

We would also like to point out that a connection between the non-perturbatively modified supersymmetry equation \eqref{susy1bD7} and a would-be ten-dimensional description of the supersymmetric AdS$_4$ solution of \cite{Kachru:2003aw} was attempted in \cite{Koerber:2007xk,Koerber:2008sx}. However, this was done by using smeared instatons, that is, roughly speaking, by replacing the NP current $\delta^{2} [\Sigma_4]$ in \eqref{eq:local-gauge} by a suitably normalized two-form proportional to $\im \Psi_1\sim J $, allowing the authors to keep the original $SU(3)$ structure. In the rest of the paper, we will pursue the same goal without relying on any smearing process (and thus going beyond $SU(3)$ structure).  

Equation \eqref{susy1b} also has intriguing implications for the geometric transition of D5 branes in supersymmetric AdS compactifications. This equation implies that the product of $\mu$ and the imaginary part of the zero-form component of $\Psi_1$ is zero. For D5-brane-type Killing spinors, this rules out AdS solutions, except for rigid SU(2) structure \cite{Petrini:2013ika}. Our results imply that this happens even in the presence of non-perturbative effects. In particular, this rules out the possibility of constructing supersymmetric AdS solutions whose internal space contains a Maldacena-N\'u\~nez throat \cite{Chamseddine:1997nm, Maldacena:2000yy}, or other throats with D5-brane-type Killing spinors such as the baryonic branch \cite{Butti:2004pk}.

Moreover, for throats with D3-type supersymmetry and NP effects on D5-branes wrapping a collapsed 2-cycle, such as \cite{Klebanov:2000hb}, supersymmetry requires to keep the cycle at zero size. For AdS compactifications, the four-form component of \eqref{susy1b} implies that the only way of keeping the 2-cycle trivial is  to have an $H_3$ flux and an SU(2) structure. The resulting  geometry would not have an SU(3) structure, so it would be different from the one in \cite{Klebanov:2000hb}.

\section{KKLT and dynamic $SU(2)$ structure}
\label{sec:SU2}

In the most common $\N=1$ compactifications of  type II string theory, the structure group of the internal manifold is taken to be $SU(3)$. Then there exists a single globally defined spinor, and the geometry is characterized by $J$ and $\Omega$.  However, according to the discussion of the previous section, this is too restrictive when NP contributions coming from D7-branes gaugino condensation are included.

As briefly discussed in section \ref{sec:GG}, the most general solution has a so-called dynamic $SU(2)$ structure determined by two  chiral spinors $\eta^{1,2}$, whose relative orientation is characterized by a position-dependent angle $\vphi$ such that\footnote{In principle there could be an extra phase in \eqref{orientation} distinguishing between D3/D7 type supersymmetry or D5 type but, as   already argued, an AdS$_4$ vacuum rules out the latter option.}
\begin{equation}
    \eta^{2\,\dagger}_+ \eta_+^1 = i \cos \vphi \ , \quad  \eta^{2\,\dagger}_- \gamma_m \eta_+^1 = i \sin \varphi \, \Theta_m \ .
    \label{orientation}
\end{equation}
 The geometry is then characterized as a mixture between the two 
 $SU(3)$ structures associated to each of the spinors\footnote{Note that if there are points where  $\sin \vphi=0$, the two SU(3) structures coincide at those points, and thus one cannot define a global $SU(2)$ structure on the tangent bundle. In that sense, the terminology ``dynamic SU(2)'' structure is not the most accurate, but we stick to it as it is more intuitive than the precise  terminology: SU$(3)\times $ SU$(3)\subset $ O$(6,6)$ structure.   }.  The invariant forms are the 1-form $\Theta$ and a real and a complex 2-form $J_2$ and $\Omega_2$ satisfying
\begin{equation}
    J_2 \wedge \Omega_2 = \Omega_2 \wedge \Omega_2 = 0 \ , \ 
    J_2 \wedge J_2 = \frac{1}{2} \Omega_2 \wedge \overline{\Omega}_2 \ , \ \imath_\Theta \Omega_2 = \imath_\Theta J_2 = 0.
\end{equation}
Schematically, we can think about this in the following way: the 2-forms play the role of the K\"{a}hler and holomorphic forms on four-dimensional subspaces, while $\Theta$ points holomorphically in the normal direction. This appears to be perfectly suited for our goal, as we will identify the sub-manifold corresponding to this four-dimensional subspace (at a particular point in the normal direction) with the 4-cycle $\Sigma_4$ wrapped by the D7-branes. The pure spinors are given by\footnote{There is a choice involved in righting down these expressions related to an SO(3) rotation in the space of the real 2-forms $\{J_2, \re \Omega_2, \im \Omega_2\}$ \cite{Heidenreich:2011ez}.}
\begin{eqnarray}
\Psi_+ &=&
e^{-i J_1} \wedge \left[
\cos \varphi \left(1-\frac{1}{2}J_2^2\right) + \sin \varphi \,  \im \Omega_2 - i J_2
\right]
,\\
\Psi_- &=& \Theta \wedge \left[
\sin \varphi \left(1-\frac{1}{2}J_2^2\right) - 
\cos \varphi \, \im \Omega_2 + i \re \Omega_2
\right]
,
\end{eqnarray}
with $J_1 \equiv \frac{i}{2} \Theta \wedge \overline{\Theta}$. 
 In our conventions the 1-form is normalized such that $h^{-1}(\Theta,\bar{\Theta}) = 2$.
We immediately see that the four-form component of $\re \Psi_+$ has the right form to be the calibration for $\Sigma_4$.  

\subsection*{The $\mathbb{P}^2$ example}

An interesting solution with this structure group was found in \cite{Heidenreich:2010ad}. The solution describes a configuration of  D7-branes on top of  O7-planes such that the D7 charge cancels locally. All of them   wrap the $\mathbb{P}^2$ on the  resolution of $\mathbb{C}^3/\mathbb{Z}_3$. Here we provide the elements of their calculations relevant for our purposes, and refer the reader interested in further details to the original paper. The orbifold action is given by the identification\beq
z^i \sim e^{2\pi i/3}z^i \ ,
\eeq
so this (non-compact) manifold can be described locally in terms of $\mathbb{Z}_3$-invariant coordinates $z$ and $u^{1,2}$, where
\beq
u^{1,2}=\frac{z^{1,2}}{z^3} \ , \quad z=\frac13 (z^3)^2 \ .
\eeq
The squared radius is given by 
\begin{equation} \label{rho}
    \rho^2 = (3|z|)^{2/3}\left(1+|u_1|^2+|u_2|^2\right).
\end{equation}
In the resolved phase, the holomorphic 4-cycle at the tip of the cone is defined by the equation $z=0$. 

An Ansatz for the ``would-be K\"{a}hler form'' is 
\begin{equation}
    J \equiv J_1 + J_2 \equiv e^{-4L_1(\rho)} j_1 + e^{2L_2(\rho)} j_2 \ , \ 
    j_1 = \frac{i\rho^2}{2} \der \rho^2 \wedge \derb \rho^2 \ , \ 
    j_2 = \frac{i}{2\rho^{2}} \left(\der \derb \rho^2 - \frac{\der\rho^2 \wedge \derb \rho^2}{\rho^2}\right).
    \label{Jbeyond}
\end{equation}
Here, $j_2$ and $j_1$ are respectively parallel and normal to the four-dimensional base. In the warped-CY limit, the Einstein frame versions of the functions $L_{1,2}$ satisfy $e^{2L_1^E} \sim e^{2L_2^E} \sim (r^2 + r_0^2)^{1/3}$, where $r = \rho^3/3$ and $r_0$ is the resolution scale, but this does not need to happen on the final geometry that contains all the desired ingredients.

We are particularly interested in two components of the supersymmetry equations, the 3-form component of \eqref{susy2}
and the 2-form component of \eqref{susy1bD7}\footnote{We do not expect this equation to get modified by the gaugino condensate, as it does not contain any contribution related to the cosmological constant term or to the  fluxes. This is consistent with the analysis of \cite{Dymarsky:2010mf}.}:
\begin{eqnarray}
    d\left[e^{3A-\phi}\sin \vphi \, \Theta\right]
    &=& 2 i \mu e^{2A-\phi}\left(\cos \vphi \, J_1 + J_2\right) -2i 
    \<S\> \delta^2[z=0], \label{susyD72} \\
    d\left[ e^{2A-\phi}\left(\cos \vphi \, J_1 + J_2\right)\right] &=& 0. 
    \label{susyD73}
\end{eqnarray}
Note that the localized term was omitted in \cite{Heidenreich:2010ad} and therefore the solution  was valid in regions away from where the NP dynamics take place. We now massage the two equations above for later use; the expression we are interested in is\begin{equation} \label{eq:ugly}
d[e^{2A-\phi}( J_1+J_2)]= e^{3A-\phi} \sin \varphi \ d\!\left(\dfrac{1-\cos\varphi}{	e^A \sin\varphi}\right)\wedge J_1 -2e^{A-\phi}\dfrac{1-\cos\varphi}{\sin\varphi}J_2\wedge \text{Re}(\mu\bar{\Theta} ) 
\end{equation}

The coordinate choice introduced above is particularly useful in describing the region near the four-cycle at $z=0$, which can be approached by scaling $z$ and $u_{1,2}$ with a small factor of $\vep$ and keeping the terms which are first-order in $\vep$. Thus, in this region one has 
\begin{equation} 
    r =\frac13 \rho^3\approx |z| \ , \ j_1 = \frac{i}{2} dz \wedge d\overline{z} \ , \  j_2 = \frac{i}{2} du^a \wedge d\overline{u}^a \ , \ \Theta = e^{-2L_1(r) + i \theta(r)}\, r \, \frac{dz}{z},
    \label{nearD7}
\end{equation}
with $a=1,2$. Crucially, all functions depend only on $r$, which parametrizes the distance to $\Sigma_4$, and the approximation is valid for $r \ll r_0$.  Now, in this region $ d(f(r)j_1)=0$, so  \eqref{susyD73} implies $d(e^{2A-\phi}J_2)=0$, or equivalently $2A-\phi+2L_2\equiv 2 l_2$  \ is constant. Implementing this in \eqref{eq:ugly} we find that   $\mu \approx 0$ in the vicinity of $\Sigma_4$. This means  that in this region the cosmological constant is only relevant at higher order. 
Thus, Eq. \eqref{susyD72} (which contains the NP contribution) is satisfied in this regime if and only if\footnote{This clarifies the relation between  \cite{Heidenreich:2010ad}, where this constant was named $c_1$, and \cite{Dymarsky:2010mf} (see also section 6.2 of \cite{Koerber:2007xk,Koerber:2008sx}). Indeed, $c_1$ is seen as the on-shell value for $W_{\mathrm{NP}}$, in other words, the gaugino condensate.} 
  \begin{equation}
    e^{3A(r)-\phi(r)-2L_1(r)+i\theta(r)} \sin \vphi \,r = \mathrm{constant} = -\frac{1}{\pi}\<S\>,
    \label{c1}
\end{equation}
where we have used the fact that $\delta^2[z=0] = \pi^{-1}\der \derb \, \re \log(z)$. 

\medskip 

The   way to extend the definition of the 1-form $\Theta$ outside this region is  \cite{Heidenreich:2010ad}
\begin{equation}
    \Theta = e^{-2L_1(r) + i \theta(r)}\, r \, \frac{dz}{z} \to e^{-2L_1(r) + i \theta(r)} \frac{\der r^2}{r},
    \label{beyond}
\end{equation}
with $r=\rho^3/3$ depending now on $z$ as well as on  $u^a$ by \eqref{rho}. Away from the D7-branes we can ignore the $\delta$ term and the 1-form  satisfies 
\begin{equation}
    d\left[e^{3A-\phi}\sin \vphi \, \Theta\right]
    = 2 i \mu e^{2A-\phi}\left(\cos \vphi \, J_1 + J_2\right).
\end{equation}
Using \eqref{Jbeyond}, together with the orthogonality of $j_1$ and $j_2$ one finds
\begin{equation}
    e^{3A(r)-\phi(r)-2L_1(r)+i\theta(r)} \sin \vphi \,r = \frac{\mu}{3} e^{2A(r)-\phi (r) +2 L_2(r)}. 
\end{equation}
If we want this solution to match the near-brane behavior, where the r.h.s. is constant and  \eqref{c1} should be satisfied, we should have 
\begin{equation}
    -\frac{1}{\pi}\<S\> = \frac{\mu}{3}e^{2l_2}.
    \label{matching}
\end{equation}
Note that this implies that the phase of $\langle S \rangle$ is the same as that of $\mu$, which confirms that in equation \eqref{susy1bD7} the cycle wrapped by the D7 branes cannot be trivialized. 

\subsection*{The KKLT AdS vacuum from 10D solution matching}

The matching condition \eqref{matching} is a crucial relation that encodes the stabilization of moduli from the 10D perspective. Let us consider its implications carefully.

First, recall that the volume of the wrapped cycle $\Sigma_4$ is given by the (imaginary part of the) K\"ahler modulus $\sigma$ 
\begin{equation}
    \mathrm{Vol}(\Sigma_4) = \sigma.
\end{equation}
On one hand,  $\mu$ is, by definition, related to the on-shell value of the full superpotential by
\begin{equation}
    \mu = e^{K/2} W_{\mathrm{KKLT}}= \frac{W_0 + W_{\mathrm{NP}}}{(2\sigma)^{3/2}}
\end{equation}
where we have used the expression for the K\"ahler potential $K = -3 \log (2\sigma)$ and, as explained before, $W_0$ is the flux contribution to the superpotential, given by the GCG generalization of the GVW superpotential which depends on the complex structure and axion-dilaton moduli, evaluated at the minimum. Moreover, the non-perturbative superpotential is proportional to the condensate:
\begin{equation}
    W_{\mathrm{NP}} = N_c \<S\>=\frac{2 \pi}{a} \<S\> ,
\end{equation}
where $a$ is defined in \eqref{WnpKKLT}. This expression is valid for a pure-glue gauge theory, otherwise the overall constant would be different.

On the other hand, $e^{2l_2}$ is not just any integration constant: it is related to the volume of the base of the resolved cone (the four-cycle $\Sigma_4$). More precisely, it is easy to see that $\frac{\pi^2}{2}e^{4l_2}$ measures the volume of the resolved $\mathbb{P}^2$, and is thus directly related to $\sigma$. For our purposes, it is sufficient to note that their relation is   
\begin{equation} \label{VolSigma}
     e^{2l_2} = \dfrac{ \sqrt{2\sigma}}{k}, 
\end{equation}
for some proportionality constant $k$ that depends on the details of the IR solution\footnote{The precise expression is 
\begin{equation}
k^2 = \int_{\Sigma_4} e^{\phi-4A}\cos \varphi \, j_2^2 = \pi^2 \, (e^{\phi-4A}\cos \varphi)|_{\Sigma_4}.
\end{equation}
In the last equality we have used the fact that, within our Ansatz, all scalar functions are constant along the cycle.}. Combining all this, we find that the matching condition \eqref{matching} written in terms of $\sigma$ is

\begin{equation}
    - \frac{3ka}{\pi^2} W_{\mathrm{NP}} = \frac{W_0 +W_{\mathrm{NP}}}{\sigma}.
\end{equation}
This is very similar to the F-flatness condition $D W_{4D} =0$ one obtains in the 4D EFT. Thus, the K\"{a}hler modulus is stabilized at a value $\sigma_*$ such that 
\begin{equation} \label{W0us}
    \boxed{W_0 = -A e^{-a \sigma_*} \left(1+\frac{3k}{\pi^2} \, a \, \sigma_*\right)}
\end{equation}
This is the ten-dimensional equation that is necessary in order to solve the supersymmetry condition \eqref{susyD72}. Up to a numerical factor, this equation agrees with the bottom-up EFT relation that was used in the KKLT construction, \eqref{KKLTminumum},  thus confirming the validity of the latter.
\newline

To recap, equation \eqref{W0us} arises as a requirement for matching the supergravity solution in the region very close the D7-branes, where the source term $\<S\> \delta^{2}[\Sigma_4]$  dominates over the cosmological constant contribution, with the solution in the region  away from the D7 branes, where the latter dominates over the former. The highly non-trivial matching between the 10D calculation performed here and the EFT F-flatness condition \eqref{KKLTminumum} provides support for the validity of the supersymmetric AdS vacuum: on one hand, the EFT appears indeed to capture the key features of the 10D physics properly and, on the other hand, an internal manifold with dynamical SU(2) structure appears to have the correct geometric properties to describe the  supersymmetric AdS$_4$ solution induced by fluxes and non-perturbative contributions coming from gaugino condensates.

\subsection*{Further comments}

\begin{itemize}
    \item Here we focused only on a subset of the supersymmetry conditions but, of course, all of them need to be satisfied. In \cite{Heidenreich:2010ad}, a  solution to the full system was obtained only in the region close to the D7-branes (but for $|z|>0$), where the effect of the curvature term can safely be ignored. The solution involves several integration constants, whose connection with clear physical boundary conditions is hard to obtain. Furthermore,  it is worth pointing out that some issues arise when analyzing the behavior very close to the D7-branes. 

\item As discussed above, the angle $\vphi$ is a function of the distance to the branes, and the structure is of the dynamic $SU(2)$ type. However, as one goes away from the branes, $\vphi$ approaches zero asymptotically. Thus, at large distances the structure is approximately $SU(3)$, but the presence of a small cosmological constant remains. This indicates that the description  proposed in \cite{Koerber:2007xk,Koerber:2008sx} in terms of smeared instantons can capture the important physics as seen from a distant observer. Also, the fact that $\mu$ and the deformation parameter $\vphi$ are small far away from the region where NP effects are large could allow for a perturbative treatment. 
           
  \item  It has been pointed out that the localized gaugino condensate acts as a source for the IASD components of the 3-form flux  $G_3$ \cite{Baumann:2010sx,Heidenreich:2010ad,Dymarsky:2010mf}. The existence of this flux  is related to the stabilized cycle being kept at finite size, as we now explain: we start by recalling that in the well known Klebanov-Strassler solution \cite{Klebanov:2000hb} the allowed fluxes are ISD.  
  This is a consequence of \eqref{susy3}, whose 3-form component in a conformal-CY compactification with $\mu=0$   is nothing else but the ISD condition \begin{equation}\label{eq:ISDfluxes}
*_6 G_3=iG_3\quad \text{ where } \quad G_3=F_3+ie^{-\phi}H_3 \ .  
  \end{equation} 
   In an AdS compactification,  $\mu\neq 0$, and   the self-duality condition no longer holds; the solution must contain IASD fluxes  whose strength is proportional to the new contribution in \eqref{susy3}. Therefore, both the finiteness of the cycle and the presence of IASD flux are consequences of having an AdS compactification. 
   
 We  can also generalize  the previous  argument to more general solutions. Again, we start from   \eqref{susy3} with $\mu=0$, which can be recast as an ISD condition for the more general complex ``flux'' polyform \cite{Lust:2008zd} 
 \begin{equation}   \label{Ggen}
     \tilde{*}_6 G =   i G \, , \quad {\rm where}   \quad  G = F + i e^{-4A} d_H \left(e^{4A - \phi} \re \Psi_1\right) \ , \quad  \tilde{*}_6\equiv - *_6 \alpha
    \end{equation}
and $\alpha$ is defined in \eqref{Mukai}. Note that this equation describes the Maldacena-N\'u\~nez \cite{Chamseddine:1997nm, Maldacena:2000yy} and baryonic-branch \cite{Butti:2004pk} solutions.
 Turning on a cosmological constant, \eqref{susy3} is modified to
   \beq \label{ISDmu}
   (1+ i \tilde{*}_6  ) G =  
    3 i  e^{-A-\phi}\bar{ \mu} \Psi_2\  .
    \eeq
  Following the same logic, we see that the finite stabilization of cycles by NP effects will always come together with generalized IASD fluxes $G$, whose  amplitude is tied to $\mu$.

\item 
Furthermore, note that Eq. \eqref{susy1} might not be the only supersymmetry condition modified by the backreaction of the localized strong-coupling effects. It is known that RR (and NSNS) background fluxes can generically produce a mass for the gaugino  on the branes. Flipping this argument around, one finds that the gaugino condensate should also act as a source for the fluxes \cite{Baumann:2010sx}, which would generate a new localized contribution to \eqref{susy3}. So far, the exact form of this new term has only been worked out for D7-branes in an SU(3)-structure background with a constant warp factor \cite{Dymarsky:2010mf}. This was seen to produce an additional localized ISD piece of type (0,3) in $G_3$. The general form of this new contribution in the GCG language will be addressed in detail in a more technical companion paper \cite{worldvolume}.  
\end{itemize}

\section{Discussion}

We have analyzed the backreacted geometry of NP effects, based on the quantum corrected supersymmetry equation \eqref{susy1b}. We have shown that for supersymmetric AdS$_4$ vacua with D7-branes undergoing gaugino condensation, this equation implies that the cycle wrapped by the branes stays at a finite size, in contrast with the usual geometric transition that takes place in Minkowski space.  Our approach should be applicable to any AdS compactification  involving  stacks of D$p$-branes undergoing gaugino condensation.  

This is the ten dimensional description of   K\"ahler moduli stabilization in type IIB string compactifications \`a la KKLT.  Furthermore, this works at the quantitative level. Indeed, we have found that the matching of the supergravity solution near the D7-branes (where the fields sourced by the gaugino condensate, found from equation \eqref{susy1bD7}, are stronger than the cosmological constant) with the solution far from these branes produces a relation, \eqref{W0us}, between the size of the four-cycle and the cosmological constant. This top-down result matches the bottom-up relation obtained by an EFT analysis in \cite{Kachru:2003aw}.

All our analysis is based on the supersymmetry equation \eqref{susy1b}.  It is worth pointing out that \cite{Dymarsky:2010mf} argued that the third susy equation \eqref{susy3} should be modified as well by the addition of a localized term. We leave the full set of modified equations, as well as a more detailed analysis of the background, for a future publication.  

Note that in this paper we have only matched certain limits of the full solution describing D7 branes with gaugino condensation. It would be very interesting to find the full solution. In particular, it would help to understand whether the naked singularity close to the D7-branes found in \cite{Heidenreich:2010ad}  is unavoidable and, if so, how it can be resolved in string theory.

One of our assumptions has been that D7-brane gaugino condensation happens 
and that it gives rise to a  supersymmetric AdS$_4$ vacuum. It would be very interesting to obtain a deeper field-theoretical understanding of this process, and establish whether and under what circumstances this happens when the worldvolume of the D7 branes has an  AdS$_4$ factor. 

Our top-down result matches the main features of the KKLT AdS vacuum obtained from the effective field theory of the K\"ahler moduli. However, this does not necessarily mean that this effective field theory is the appropriate one, and/or whether there are missing light modes, as suggested both by swampland arguments  \cite{Lust:2019zwm}, and by an explicit computation of the Laplacian in the warped throat \cite{Blumenhagen:2019qcg}. Furthermore, taking into account  the warp factor in the EFT for the complex structure moduli was shown to lead to a potential with a very different behavior than the constant-warp-factor potential \cite{Bena:2018fqc}. The minima of the two potentials are the same, but their behaviors near the origin are completely different. In particular, the naive constant-warp-factor potential fails to capture the runaway behavior that is produced by the addition of  anti-D3-branes.

Furthermore, as highlighted in \cite{Moritz:2017xto,Gautason:2018gln,Hamada:2018qef,Kallosh:2019oxv,Hamada:2019ack,Gautason:2019jwq,Carta:2019rhx}, any attempt to uplift the cosmological constant via the inclusion of some localized objects (such as anti-D3 branes) is highly dependent on the details of the gaugino condensation. Now that we understand (albeit by patches) the features of the 10D solution sourced by D7 branes with a condensate, it would be very interesting to construct a more general 10D geometry (and possibly an effective 4D EFT) where one takes into account also the effects of the anti-D3 branes.

We also argued that the AdS$_4$ supersymmetry equations rule out  D5-brane-type supersymmetry (except for rigid SU(2) structure), even in the presence of non-perturbative effects. As we have explained, this rules out supersymmetric AdS compactifications with Maldacena-N\'u\~nez or baryonic-branch throats. Throats with D3-brane-type supersymmetry, instead, can be embedded in AdS compactifications only if the D5-brane developing a gaugino condensate wraps a collapsed cycle. This can only be achieved if the internal manifold has SU(2) structure, so solutions such as  the Klebanov-Strassler throat would have to be modified into a manifold with this structure group.

We would also like to stress that the procedure we  used to match the near- and far-D7 brane regions and to obtain the 10D version of moduli stabilization appears to be more general and can also be applied to other D-branes wrapping calibrated cycles. For type IIB compactifications the only such branes are D5 branes wrapping two-cycles but, as we have seen above, these are already ruled out in AdS  compactifications. However, in type IIA compactifications our procedure could in principle be used to study the backreaction of D-branes with gaugino condensation and to obtain similar top-down equations governing the moduli of supersymmetric AdS vacua.

\label{sec:conclusions}

\appendix

\acknowledgments
We thank Eduardo Garc\'{i}a-Valdecasas, Severin L\"ust, Luca Martucci, Liam McAllister, Jakob Moritz, Radu T\u{a}tar, Alessandro Tomasiello and \'{A}ngel Uranga  for interesting comments and valuable insights. N.K. is grateful to the Institut de Physique Th\'eorique for kind hospitality while working on this project. 
This work was partially supported
by the ERC Consolidator Grant 772408-Stringlandscape,  the ANR grant Black-dS-String ANR-16-CE31-0004-01 and the John Templeton Foundation grant 61169. The work of N.K. has additionally been supported  by the National Agency for the Promotion of Science and Technology of Argentina (ANPCyT-FONCyT) Grants PICT-2015-1525, PICT-2017-1647 and the ERC Starting Grant 639729-preQFT.

\newpage

\bibliographystyle{JHEP}
\bibliography{refs}

\end{document}